\documentclass[a4paper]{jpconf}
\usepackage{iopams}

\usepackage{graphicx}
\usepackage{grffile}
\usepackage{color}
\usepackage{wrapfig}

\definecolor{slateblue}{rgb}{0.2,0.2,0.6}
\usepackage[colorlinks=true,linkcolor=slateblue,citecolor=slateblue,urlcolor=slateblue,pdfauthor={MOrcinha},pdftitle={TimeLagECRS}]{hyperref}

\newcommand{\ApJ}{ApJ}
\newcommand{\ApJL}{ApJL}
\newcommand{\AeA}{A\&A}
\newcommand{\PRD}{PRD}
\newcommand{\ASR}{ASR}

\newcommand{\PLB}{PLB}
\newcommand{\MNRAS}{MNRAS}

\newcommand{\AMS}{\textsf{AMS}}

\newcommand{\ie}{\textit{i.e.}}

\newcommand{\epratio}{{e\ensuremath{^{+}}}/{e\ensuremath{^{-}}}}

\newcommand{\DT}{\ensuremath{\Delta{T}}}
\newcommand{\Kappa}{\ensuremath{\boldsymbol{K}}}

\begin{document}
\title{Observation of a time lag in solar modulation of cosmic rays in the heliosphere}

\author{Miguel Orcinha$^{\,1}$, Nicola Tomassetti$^{\,2}$, Fernando Bar\~{a}o$^{\,1}$, Bruna Bertucci$^{\,2}$}
\address{$^{1}$\,Laborat{\'o}rio de Instrumenta\c{c}{\~a}o e F{\'i}sica Experimental de Part{\'i}culas, P-1000 Lisboa, Portugal}
\address{$^{2}$\,Universit{\`a} degli Studi di Perugia \& INFN-Perugia, I-06100 Perugia, Italy;}
\ead{migorc@lip.pt}

\begin{abstract}
The solar modulation effect of Galactic cosmic rays is a time-dependent phenomenon that is caused by the transport of these particles through the magnetized plasma of the heliosphere. Using a data-driven model of cosmic-ray transport in the heliosphere, in combination with a large collection of data, we report the evidence for a eight-month time lag between observations of solar activity and measurements of cosmic-ray fluxes in space. As we will discuss, this result enables us to forecast the cosmic ray flux at Earth well in advance by monitoring solar activity. We also compare our predictions with the new multi-channel measurements of cosmic rays operated by the \AMS{} experiment in space.
\end{abstract}

\section{Introduction}
New-generation experiments of cosmic ray (CR) detection have reached an unmatched level of precision
that is bringing transformative advances in astroparticle physics \cite{AmatoBlasi2017,Grenier2015}.
Along with calculations of CR propagation in the galaxy,
the interpretation of the data requires detailed modelling of the so-called \emph{solar modulation effect}.
Solar modulation is experienced by all CR particles that enter the heliosphere to reach our detectors near Earth.
Inside the heliosphere, CRs travel through a turbulent magnetized plasma,
the solar wind, which significantly reshapes their energy spectra. This effect is known to change with time,
in connection with the quasi-periodical $11$-year evolution of the
solar activity, and to provoke different effects on CR particles and antiparticles \cite{Potgieter2014}.

Observationally, an inverse relationship between solar activity (often monitored by the number of solar sunspots)
and the intensity of CRs at Earth is known for a long time. The effect of solar modulation in the low-energy CR
spectra ($E\lesssim$\,GeV) is measured by several experiments \cite{Bindi2017,ValdesGaliciaGonzalez2016}.
Solar modulation is caused by a combination of basic particle transport processes such as diffusion, convection, adiabatic cooling, or drift motion,
yet the underlying physical mechanisms and their associated parameters remain under active investigation.

Solar modulation models are dependent on two crucial factors:
(i) precise knowledge of the interstellar spectra (LIS) of CRs outside the heliosphere;
(ii) availability of time-series of CR data on different species.
Recent accomplishments from strategic space missions have enabled us to make significant progress in this field.
The entrance of Voyager-1 in the interstellar space
provided us with the very first LIS data on CR protons and electrons \cite{Stone2013,Cummings2016}.
Long-duration space experiments PAMELA (on orbit since 2006) and \AMS{} (since 2011)
have been releasing a continuous stream of monthly-resolved data on CR particles and antiparticles \cite{Adriani2013,Adriani2016,Aguilar2018ProtonHelium,Aguilar2018ElectronPositron}.

These measurements add to a large wealth of low-energy CR data collected in the last decades by space missions CRIS/ACE \cite{Wiedenbeck2009}
IMP-7/8 \cite{GarciaMunoz1997}, \emph{Ulysses} \cite{Heber2009}, and more recently EPHIN/SOHO \cite{Kuhl2016},
as well as from ground data provided continuously by the neutron monitor (NM) worldwide network \cite{Mavromichalaki2011,Steigies2015}.

In this article, we present the work done on \cite{Tomassetti2017} which presents new calculations of CR fluxes near-Earth that account for the dynamics of CR modulation in the expanding heliosphere and expand on the understanding of the correlation between solar parameters and the CR fluxes.
Using a large collection of modulated and interstellar CR data collected in space,
we have constructed a predictive and measurement-validated model of solar modulation which depends
only on direct solar-activity observables: the sunspot number (SSN) and the tilt angle of the heliospheric current sheet (HCS).

\section{Methodology}
The transport of CRs in heliosphere is described by the Parker equation for the omni-directional phase
space density  $\psi(t,p,\boldsymbol{ r})$ expressed as function of time $t$, momentum $p$, and position $\boldsymbol{r}$ \cite{Potgieter2013}:
\begin{equation}\label{Eq::TransportEquation}
  \frac{\partial \psi}{\partial t} = - (\boldsymbol{V} + \boldsymbol{v_{d}}) \cdot \nabla \psi + \nabla \cdot ({\Kappa} \cdot \nabla \psi) + \frac{1}{3}(\nabla \cdot \boldsymbol{V}) \frac{\partial \psi}{\partial \ln{p}}
\end{equation}
The various terms represent convection with a solar wind of speed $\boldsymbol{|V|}\cong$\,400\,km/s,
drift motion with average speed $\boldsymbol{v_{d}}$, spatial diffusion with tensor ${\Kappa}$,
and adiabatic momentum losses.

We simulated this physical process in a minimal 2D description using $\boldsymbol{r}=(r,\theta)$, radius and heliolatitude \cite{Bobik2012}.
In terms of diffusion coefficient, the parallel component was defined as $K_{\parallel}=  \kappa^{0} \frac{10^{22}\beta p/{\rm GeV}}{3B/B_{0}}$, in units of cm$^{2}$/s, where we have factorized an adimensional scaling factor, $\kappa^{0}$, of the order of unity.
The perpendicular component is $K_{\perp}\cong 0.02\,K_{\parallel}$ \cite{GiacaloneJokipii1999}.

\begin{wrapfigure}{O}[0mm]{0.4\textwidth}
\centering
\includegraphics[width=0.35\textwidth]{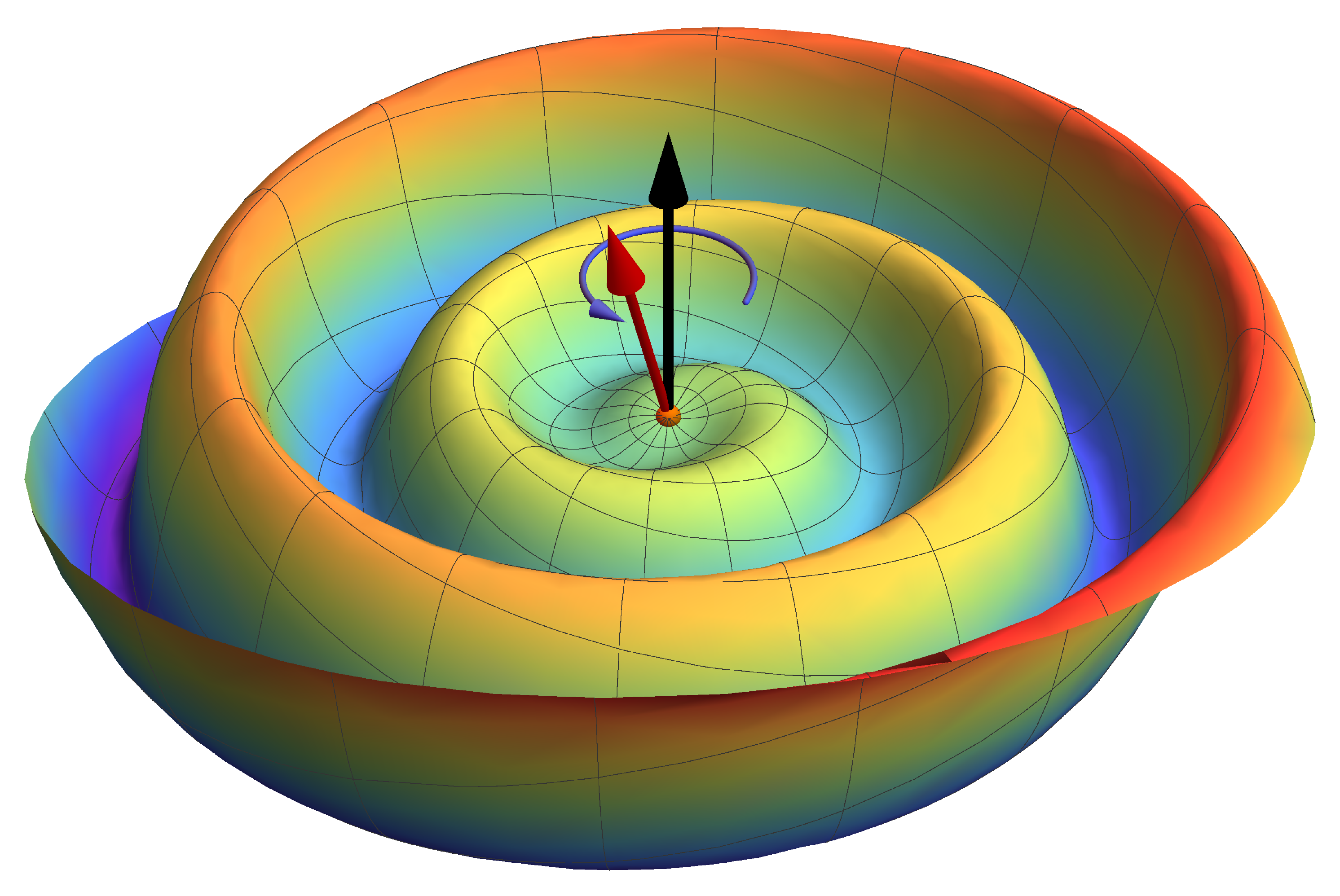}
\caption{\footnotesize{
Graphical representation of the heliospheric current sheet in a Parker solar wind model. The black arrow represents the axis of rotation and the red one the magnetic dipole direction. Tilt angle is then defined as the angle between these two vectors.
}}
\label{Fig::ccHCS}
\vspace{-2mm}
\end{wrapfigure}

The regular solar magnetic field (HMF) is modelled using the usual Parker structure,
$B=\frac{B_{0}A}{r^{2}}\sqrt{1+\Gamma}$ with $\Gamma = \left(\Omega r/V\right) \sin{\theta}$,
where $\Omega=2.866\cdot\,10^{-6}$\,rad\,s$^{-1}$ is the angular rotation of the Sun,
$B_{0}\cong$\,3.4\,nT\,AU$^{2}$ is the field intensity at $r_{0}=$\,1\,AU,
and $A=\pm\,1$ represents the magnetic polarity cycle of the Sun.
The polarity is positive (negative) when the HMF points outward (inward) in the north hemisphere.

In this model we account for gradient and curvature drift effects.
In particular, drift is important across the wavy layer of the HCS, \ie{} the surface where magnetic polarity
changes from north to south, the angular extension (and waviness) of which is described by the tilt angle $\alpha$ (see Figure \ref{Fig::ccHCS}).
The drift velocity components $v_{r}$ and $v_{\theta}$ are both proportional to $qA\frac{2\beta r p }{3B_{0}}$,
so that the sign of $\boldsymbol{{v}_{d}}$ depends on the product $qA$ \cite{Potgieter2014}.

With this setting, we compute the CR propagation from the termination shock (TS) to Earth's orbit using the stochastic differential equation approach of \cite{Kappl2016}.
This method consists of a backwards-in-time propagation of a large number of pseudo-particles from Earth to the boundaries \cite{Raath2016,Strauss2012,AlankoHuotari2007}.
For a given particle type, steady-state solutions of Eq.\,\ref{Eq::TransportEquation} ($\partial\psi/\partial{t}=0$) are then obtained by sampling.
In our model we disregard re-acceleration effects occurring at the TS or modulation in the heliosheath \cite{Langner2003,LangnerPotgieter2004}.

The LIS fluxes  $J^{\rm IS}$ were calculated using an improved model of CR acceleration and propagation \cite{Tomassetti2015TwoHalo,TomassettiDonato2015,Feng2016}, being well constrained by Voyager-1 and \AMS{} data \cite{Cummings2016,Aguilar2015Proton}.
Positron LIS relies on secondary production calculations so it has larger uncertainties \cite{Feng2016}.

\begin{wrapfigure}{O}[0mm]{0.4\textwidth}
\centering
\includegraphics[width=0.4\textwidth]{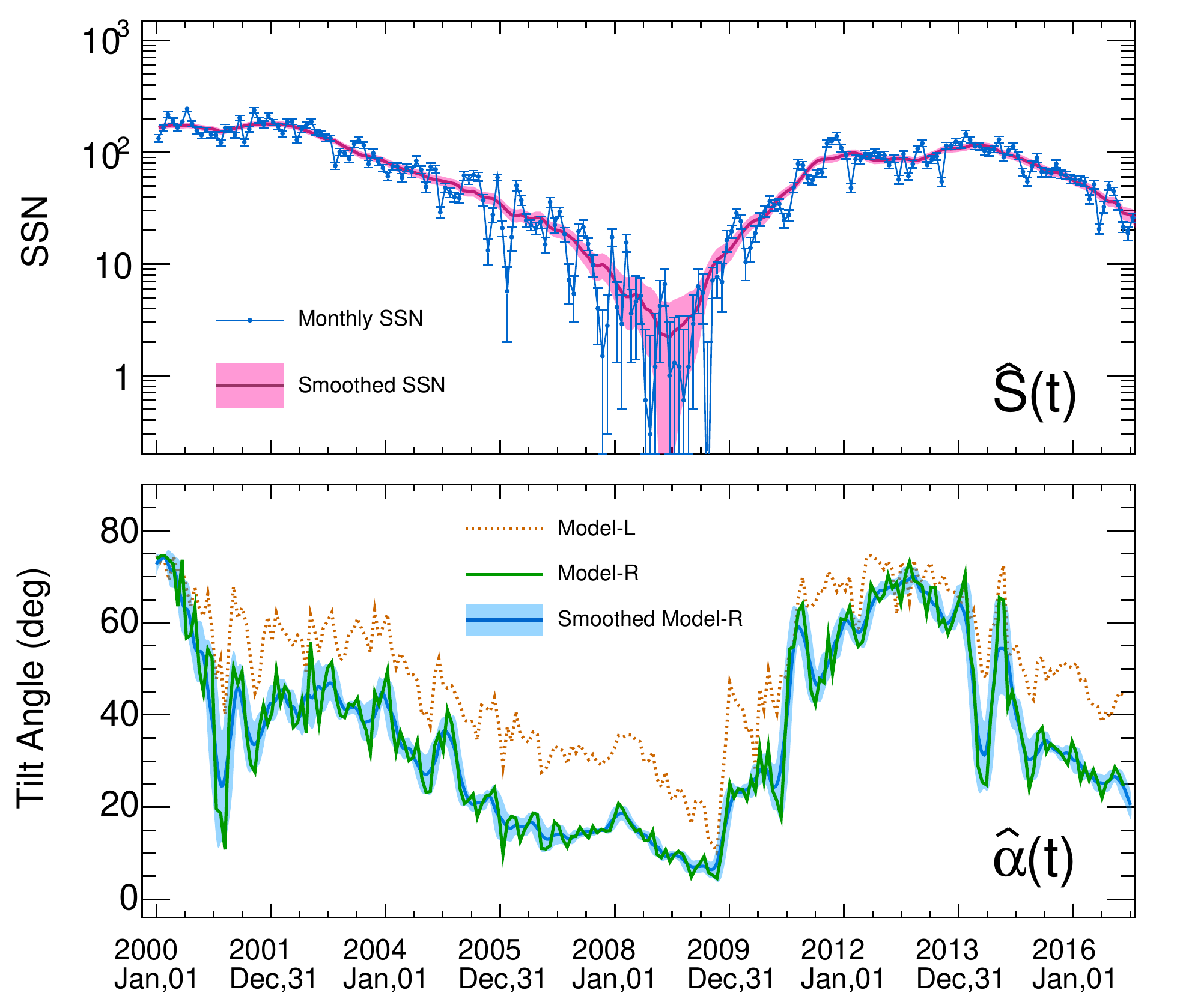}
\caption{\footnotesize{
Reconstruction of the sunspot number ($\hat{S}(t)$) and tilt-angle of the heliospheric current sheet ($\hat{\alpha}(t)$) as function of time. Figure taken from \cite{Tomassetti2017}.
}}
\label{Fig::ccSolarData}
\end{wrapfigure}

The role of positive and negative particles interchanges with polarity as can be seen in Figure \ref{Fig::ccAntimatterMatterRatios_eeplus}.
This phenomenon is due to sign-dependence of the the drift motion of particles. Depending on solar polarity CR particles and antiparticles sample different parts of the heliosphere.
Along with $qA$, the interplay of the various physics processes depends on the
levels of HCS waviness (see Figure \ref{Fig::ccHCS}) and HMF irregularities, that are here parametrized by $\alpha$ and $\kappa^{0}$.
Tilt-angle measurements $\hat{\alpha}$ are provided on 10-day basis from the \emph{Wilcox Solar Observatory}\footnote{\url{http://wso.stanford.edu}}, from which we used the improved model-$R$ \cite{Hoeksema1995,FerreiraPotgieter2004}.
A basic diagnostic for the HCS turbulence level is the manifestation
of solar sunspots \cite{Plunian2009}, so that SSNs and $\kappa^{0}$ are related each other \cite{Boschini2017,Bobik2012}.
We used the the monthly-series of SSNs provided by the \emph{Royal Observatory of Belgium}\footnote{\url{http://www.sidc.be}}.
Both $\alpha$ and SSN were smoothly interpolated in order to create the functions $\hat{\alpha}$ and $\hat{S}$
For diffusion we adopted a simple two-coefficient relation $\kappa^{0} \equiv a+b\log({\hat{S}})$.

The temporal behaviour of these quantities, shown in Fig.\,\ref{Fig::ccSolarData}
from 2000 to 2017, is at the basis of the time-dependent nature of solar modulation.
In practice, the problem is modelled under a \emph{quasi-steady} fashion, \ie, by providing a
time-series of steady-state solutions corresponding to a time-series of input parameters.
Due to the nature of the phenomenon, a finite amount of time is needed, in fact, for the properties observed in the solar corona
to be transported in the outer heliosphere by the plasma.
This motivated us to introduce a parameter \DT{} in our calculation, describing
a time-lag between the solar-activity indices of Fig.\,\ref{Fig::ccSolarData} and
the medium properties of the modulation region, \ie, the spatial region effectively sampled by CRs.
Evidence for a lag of $\sim$\,6-12 months have been previously reported in NM-based
studies \cite{MavromichalakiPetropoulos1984,Badruddin2007,Lantos2005,Nymmik2000,MishraMishra2016,Chowdhury2016}.

Our model is then specified by three free parameters only, $a$, $b$, and \DT, that we constrain using a large amount of data.
We use monthly-resolved proton data from the PAMELA experiment \cite{Adriani2013}
collected between July 2006 and January 2010,
and data from the EPHIN/SOHO space detector \cite{Kuhl2016}, yearly-resolved between 2000 and 2016.
We also include data from the BESS Polar-I (Polar-II) mission from 13 to 21 December 2004 (from 23 December 2007 to 16 January 2008) \cite{Abe2008,Abe2012}.
These measurements are given in terms of time-series of energy spectra $\hat{J}_{j,k}=\hat{J}(t_{j},E_{k})$,
where each spectrum is a \emph{snapshot} of the CR flux near-Earth at epoch $t_{j}$.
Calculations $J(t_{j}, E_{k})$  are performed using retarded functions of the physics inputs
$\alpha_{j}=\hat{\alpha}(t_{j}-\DT)$ and  $\kappa^{0}_{j} = a+b\log(\hat{S}(t_{j}-\DT))$.
We then build a global $\chi^{2}$-estimator:
\begin{equation}\label{Eq::GlobalChiSquare}
\chi^{2}(a,b,\DT) = \sum_{j,\,k} \left[ \frac{ J(t_{j}, E_{k}; a, b, \DT) - \hat{J}_{j,k} }{\sigma_{j,k} } \right]^{2}
\end{equation}

The quantity $\sigma_{j,k}$ includes experimental errors in the data and model uncertainties due to finite statistics
of the pseudo-particles simulation. The following sources of systematic uncertainties are also accounted:
(i) uncertainties in the LIS, from the constraints provided by Voyager-1 and \AMS{} data;
(ii) uncertainties on $\hat{\alpha}(t)$, from the discrepancy between $L$ and $R$ models and from the smoothing procedure;
(iii) uncertainties on $\hat{S}(t)$  from the smoothed SSN variance, see Fig.\,\ref{Fig::ccSolarData}.
The free parameters are estimated by means of standard minimization techniques.

\section{Results}

\begin{figure*}[!t]
\centering
\includegraphics[width=0.7\textwidth]{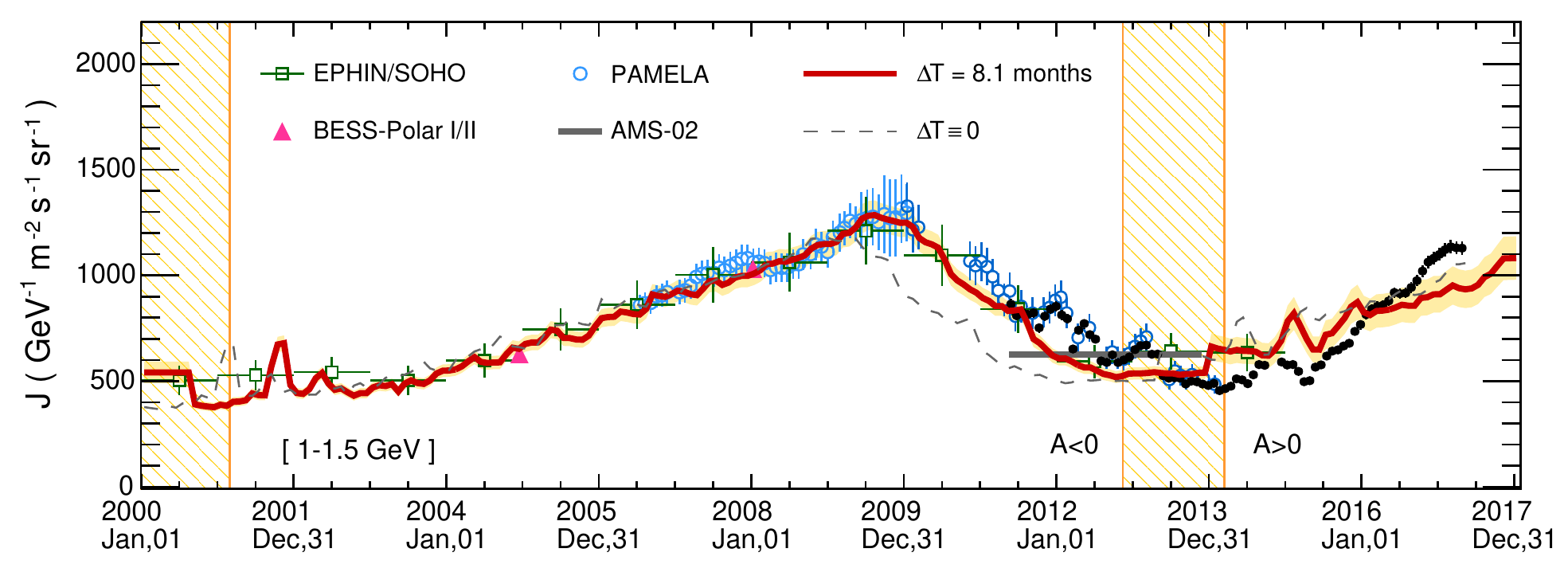}
\caption{\footnotesize{%
Time profile of the proton flux at $E=1-1.5$\,GeV. Best-fit calculations are shown as thick solid line,
along with the uncertainty band, in comparison with the data \cite{Adriani2013,Kuhl2016,Abe2008,Abe2012,Aguilar2015Proton,Aguilar2018ProtonHelium,Martucci2018Proton}. The most recent \AMS{} data\cite{Aguilar2018ProtonHelium} is shown in black and PAMELA data\cite{Martucci2018Proton} in dark blue.
The shaded bars indicate the magnetic reversals of the Sun's polarity \cite{Sun2015}.
Figure taken from \cite{Tomassetti2017}.
}}
\label{ccProtonFluxTimeProfile_PAMELA_AMS02_2018}
\vspace{-2mm}
\end{figure*}

The global fit has been performed to 3993 proton data points collected between 2000 and 2012 (in $A<0$ conditions)
at kinetic energy between 0.08 and  50\,GeV. The best-fit parameters are
$\hat{a}=$\,3.88\,$\pm$\,0.87, $\hat{b}=$\,-1.30\,$\pm$\,0.29, and $\widehat{\DT}$=8.1\,$\pm$\,0.9\,months, giving $\chi^{2}/df=2651/3990$.
The fit was repeated after fixing $\DT\equiv\,0$, \ie, under a more conventional ``unretarded'' scenario.
returning $\hat{a}=$\,3.36\,$\pm$\,0.76, $\hat{b}=$\,-1.08\,$\pm$\,0.24, and $\chi^{2}/df=$\,4979/3991.
Results are shown in Fig.\,\ref{ccProtonFluxTimeProfile_PAMELA_AMS02_2018},
illustrating the time and energy dependence of calculations in comparison with the data.
The fits give satisfactory results at all energies and epochs.
From Fig.\,\ref{ccProtonFluxTimeProfile_PAMELA_AMS02_2018}, it can be seen that the model reproduces very well
the time evolution of the proton flux, at $E=1.5$\,GeV and after the 2013 polarity reversal, the proton flux is predicted to increase with time.

It is also clear from the figure that the retarded scenario (with \DT$\equiv$\,8.1\,months, thick red line)
allows for a much better description of the time evolution of the proton flux.
Additionally, for comparison, recent \AMS{} and PAMELA data were overlaid as well even though they are not used for the fit. The prediction shows great agreement with this recent data but some fine tuning is required. Due to the change of particle drift motion during and after a magnetic reversal, we expect that a new $\DT$ has to be determined for each epoch.

Additionally, we have also performed a time-series of fits to single
energy spectra $\hat{J}_{j}$ by directly using the diffusion scaling as free parameter.
This provided time-series of 62 $\hat{\kappa}^{0}_{j}$-values corresponding to various epochs $t_{j}$,
which were done after fixing $\DT\equiv\,\widehat{\DT}=$\,8.1\,months.
Inspecting the relation between $\hat{\kappa}^{0}_{j}$ and the ``delayed'' SSN revealed a correlation coefficient of $\rho_{\DT}=-0.89$ against the $\rho_{0}=-0.66$ for the scenario of $\DT=$\,0.
Figures for both $\DT=$\,8.1\,months and $\DT=$\,0 scenarios can be found in \cite{Tomassetti2017}.

These findings explain why other authors, when proposing simple relations between $\kappa^{0}$ and SSN,
had to adopt different coefficients for descending and ascending phases \cite{Bobik2012,Boschini2017}.
Remarkably, this problem is naturally resolved in our model:
once the time-lag is properly accounted, the $\kappa^{0}$-SSN relations can be described by a unique function.

\begin{figure*}[!t]
\centering
\includegraphics[width=0.7\textwidth]{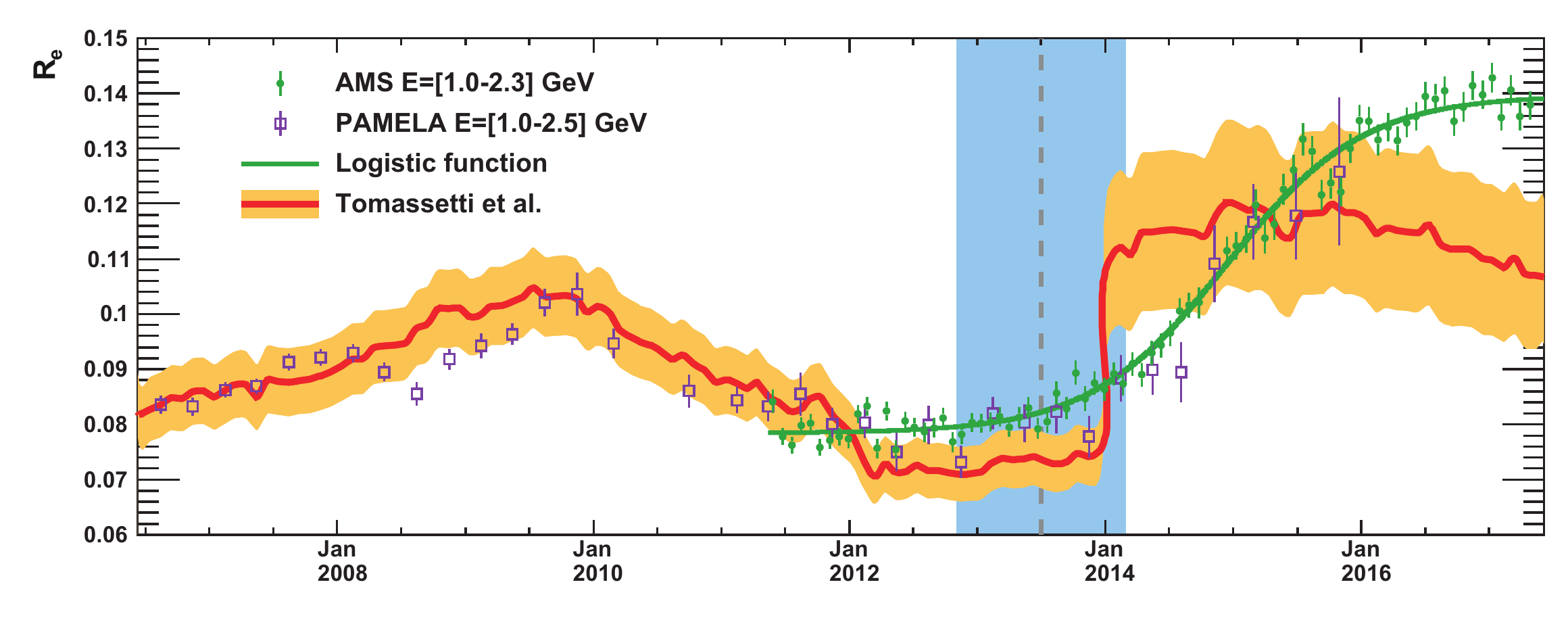}
\caption{\footnotesize{%
Time profile of the ratios $R_e$ \epratio{} at $E=1\--2.3$\,GeV for \AMS{} data\cite{Aguilar2018ElectronPositron} and at $E=1\--2.5$\,GeV for PAMELA data.
Model predictions and their corresponding uncertainties, are shown in comparisons
with the data \cite{Adriani2016,Clem2009}.
The shaded bars indicate the magnetic reversals of the Sun's polarity \cite{Sun2015}.
Figure taken from the supplementary material of \cite{Aguilar2018ElectronPositron}.
}}
\label{Fig::ccAntimatterMatterRatios_eeplus}
\vspace{-2mm}
\end{figure*}

Finally, our model was used to predict the time evolution of antimatter-to-matter ratios such as the \epratio{}.
Our calculations are shown in Fig.\,\ref{Fig::ccAntimatterMatterRatios_eeplus} for \DT=8.1\,months.
Measurements of the relative variation of the ratios are shown as reported by \AMS{} \cite{Aguilar2018ElectronPositron} and PAMELA \cite{Adriani2016}.
For the \epratio{} ratio, we note that calculations within $\DT=8.1$\,month are favoured, although the data does not permit a resolute discrimination.
Across the magnetic reversal, shown in the figure as shaded bars, a remarkable increase of the \epratio{} ratio is predicted.
It should be noted, however, that the dynamics of the transition could not be modelled during reversal, because the HMF polarity is not well defined.
Nonetheless, a rise in the \epratio{} ratio profile has been detected in both \AMS{} \cite{Aguilar2018ElectronPositron} and PAMELA data \cite{Adriani2016},
and this rise is found to occur a few months after completion of the Sun's polarity reversal.

\section{Conclusions}
Using new high-statistics measurements of CRs in space have determined the evolution of CR fluxes near Earth with unmatched time resolution.
These data allow us to perform detailed studies of the solar modulation effect and its dynamical connection with the evolving solar activity.
In this article, we have reported new calculations of CR modulation based on a simple but physically consistent numerical model
that accounts for particle diffusion, drift, convection and adiabatic cooling.
We have adopted a simple formulation where the time-dependent physics inputs of the model consist only in SSN and HCS tilt angle.
We have shown that this model reproduces well the time evolution of the Galactic proton spectra measured by \AMS{}, PAMELA, EPHIN/SOHO, and BESS experiments.
Our model is highly predictive once the correspondence between modulation parameters and solar-activity indices is established.
Our study revealed an interesting aspect of the dynamics of CR modulation in the expanding wind,
that is, the presence of time-lag \DT{} between solar data and the condition of the heliosphere.

Using a large ensemble of CR proton data we found $\DT=$\,8.1\,$\pm$\,0.9\,month, which is in agreement with
basic expectations and with recent NM based analysis \cite{AslamBadruddin2015,MishraMishra2016,Chowdhury2016}.
An interesting consequence of this result is that the galactic CR flux at Earth can be predicted, at any epoch $t$,
using solar-activity indices observed at the time $t-\DT$. This result is of great interest for real-time
space weather forecast, which is an important concern for human space-flight.

In our results, the parameter \DT{} has been determined using CR protons during negative polarity.
This parameter has to be viewed as an effective quantity representing
the average of several CR trajectories in the heliosphere during $A<0$ conditions.
Further elaborations may include the use of NM data, for larger observation periods, or the accounting for
a latitudinal dependence in the wind profile or in the diffusion coefficient.
Since CR particles and antiparticles sample different regions of the heliosphere, we expect slightly different
time-lags, $\DT_{\pm}$ depending on the sign of $qA$ (and in particular, $\DT_{-}\lesssim\,\DT_{+}$).
With the precision of the existing data, we were unable to test this hypothesis.
A detailed re-analysis of our model, in this direction, will be possible after the release
of monthly-resolved data from \AMS{} on CR particle and antiparticle fluxes.

\ack{
  We thank our colleagues of the \AMS{} Collaboration for valuable discussions.
  MO acknowledges support from FCT - Portugal, \emph{Funda\c{c}\~{a}o para a Ci\^{e}ncia e a Tecnologia}, under grant SFRH/BD/104462/2014.
  NT and BB acknowledge the European Commission for support under the H2020-MSCA-IF-2015 action, grant No.707543
  MAtISSE -- \emph{Multichannel investigation of solar modulation effects in Galactic cosmic rays.
}

\section*{References}
\bibliographystyle{iopart-num}

\end{document}